# Screening COVID-19 Based on CT/CXR Images & Building a Publicly Available CT-scan Dataset of COVID-19

Maryam Dialameh[1], Ali Hamzeh[2], Hossein Rahmani[3], Amir Reza Radmard[4], and Safoura Dialameh[5]

**Abstract**

The rapid outbreak of COVID-19 threatens humans' life all around the world. Due to insufficient diagnostic infrastructures, developing an accurate, efficient, inexpensive, and quick diagnostic tool is of great importance. As chest radiography, such as chest X-ray (CXR) and CT computed tomography (CT), is a possible way for screening COVID-19, developing an automatic image classification tool is immensely helpful for detecting the patients with COVID-19. To date, researchers have proposed several different screening methods; however, none of them could achieve a reliable and highly sensitive performance yet. The main drawbacks of current methods are the lack of having enough training data, low generalization performance, and a high rate of false-positive detection. To tackle such limitations, this study firstly builds a large-size publicly available CT-scan dataset, consisting of more than 13k CT-images of more than 1000 individuals, in which 8k images are taken from 500 patients infected with COVID-19. Secondly, we propose a deep learning model for screening COVID-19 using our proposed CT dataset and report the baseline results. Finally, we extend the proposed CT model for screening COVID-19 from CXR images using a transfer learning approach. The experimental results show that the proposed CT and CXR methods achieve the AUC scores of 0.886 and 0.984 respectively. The proposed CT-scan dataset is available here: https://github.com/m2dgithub/CT-COV19.git

[1] M. Dialameh is with the Department of Computer Science, Shiraz University, Shiraz, IRAN, e-mail: 4tiamo4@gmail.com.
[2] A. Hamzeh is with the Department of Computer Science, Shiraz University, Shiraz, IRAN, e-mail: s_ali@cse.shirazu.ac.ir.
[3] H. Rahmani is with the School of Computing and Communications, Lancaster University, UK, e-mail: h.rahmani@lancaster.ac.uk.
[4] A. R. Radmard is with the Department of Radiology, Tehran University of Medical Sciences, Tehran, Iran, e-mail: amir.radmard@gmail.com.
[5] S. Dialameh is with the School of Paramedical Sciences, Bushehr University of Medical Sciences, Bushehr, Iran, e-mail: s.diyalameh@bpums.ac.ir.

# 1. Introduction

COVID-19 pandemic has seriously affected people's lives, governments, and the economy all around the world. The early and accurate diagnosis of COVID-19 plays a vital role in both disease treatment and isolation. Current clinical methods such as real-time polymerase chain reaction (RT-PCR) [43] are insufficient in terms of both availability and accuracy, e.g., a high rate of false-positive and low sensitivity [42, 10]. Apart from that, such methods are costly in terms of safety, human resources, and financial burdens [39]. Therefore, governments as well as hospitals cannot effectively tackle such a devastating pandemic.

Diagnosis by analyzing chest radiography, such as chest X-ray (CXR) and CT computed tomography (CT), is another possibility for screening COVID-19 [3]; however, analyzing such images by experts/specialists is tedious, time-consuming [10], and possibly not accurate. Moreover, the early detection of COVID-19 by only visually monitoring chest imaging data would be almost impossible, as the patterns of the infections in lungs cannot be easily detected by human visual systems during the first days of the disease. Therefore, developing an automatic and accurate image processing tool seems promising to address the aforementioned problems. Due to the great achievements of deep learning methods in a variety of domains [4, 22, 23, 33], researchers have recently proposed several interesting deep learning approaches for automatically classifying chest images as positive/negative COVID-19 [11, 28, 40, 45]. Based on the type of data, these methods could be classified into two main categories: CXR and CT based models. Detailed information about datasets and methods can be explored in [1, 15, 27, 36, 38].

One of the main limitations of the current deep learning models is the size of datasets being used for learning the models, i.e., the number of publicly available training samples used for optimizing parameters. To the best of our knowledge, current models have been trained on small-sized datasets, mostly due to privacy concerns and the unavailability of COVID-19 CXR/CT images [26, 30]. Consequently, it does lead to a lower generalization ability of the trained models [6]. Age and regional diversities are two other important factors [7, 21, 31], playing an essential role in the generalization of the learned models, preventing the models from the danger of overfitting. As elderly people are at a higher risk to be infected by coronavirus, current public datasets of COVID-19 are often biased towards older people, resulting in a less chance of generalization for younger people. Additionally, people of different regional backgrounds are not necessarily common in their lungs functioning, respiration abilities, and several other breathing factors [9]. Hence, if a dataset is built based on the samples provided by one hospital/city, the models learned on such datasets would be very likely to bias toward that specific region. Finally, the quality of images should be preserved at an acceptable level/resolution, while this is not the case in some of the existing publicly available COVID-19 datasets.

To overcome the aforementioned limitations, this study builds a publicly available CT-scan dataset of COVID-19, consisting of 13k CT-images captured from more than 1000 individuals. The images are collected from four regions with entirely different climate conditions. A wide range of age diversities has been also included, in which ages vary from 19 to 73. Additionally, images are saved in a high-level of quality. Overall, the proposed dataset, named CT-COV19, provides a reliable set of CT-scan images for the researchers to develop more accurate and general models for screening COVID-19.

This study also proposes a novel deep neural screening model, named as Deep-CT-Net, which is trained on the proposed CT-COV19 and provides a baseline result on this dataset. The carefully designed but simple deep neural network model enables the possibility of early screening symptoms of COVID-19 based on CT-images of lungs. More precisely, the proposed model benefits from a pyramidal attention layer [20], providing pixel-wise attention on high-level features and consequently enabling the whole model to effectively detect COVID-19 even when there are small clues of the disease in the lungs. Having no heavy pre-processing steps, such as lung segmentation used in [13, 18], is another advantage of the proposed method, enabling the model to detect COVID-19 patients with a lower computational time. Extensive experiments on several benchmark COVID-19 datasets show that the proposed CT-based model outperforms the state-of-the-art methods [5, 29, 41, 44, 2, 13] in both area under curve (AUC) and accuracy.

Additionally, a transfer-learning version of this model, named Deep-CXR-Net, is further developed to screen COVID-19 based on CXR images. The choice of transfer learning enables the network to learn from unlabeled CXR images and then adjusts the weights by a small set of labeled CXR images. The experimental results show a higher rate of generalization abilities compared to several related methods such as [11, 12, 45].

## 2. Results

### 2.1. CT-COV19: A public CT-scan dataset for COVID-19

Approving by institutional review board (IRB), this section describes the details of our publicly available CT-scan dataset named CT-COV19 for screening COVID-19. CT-COV19 consists of 13k CT-images of lungs, which are taken from more than 1000 randomly selected male and female individuals. Among the patients, 500 of them were infected with COVID-19. An RT-PCR test was performed to confirm their infections with COVID-19. Moreover, the age of individuals ranges from 19 to 73. Therefore, CT-COV19 is diverse in terms of both gender and age groups. The regional diversity has also been included in our dataset, as it has been collected from four different regions with diverse climates. It is worthwhile noting that the collected data are anonymous and privacy concerns are satisfied. Figure 1 provides a graphical overview of the age and gender diversities.

One of the main advantages of this dataset is the number of COVID-19 samples, which is the biggest size among the publicly available datasets of COVID-19 so far. Another important aspect of CT-COV19 is the presence of pneumonia data, providing learning algorithms the opportunity of distinguishing between infections caused by COVID-19 and other lung diseases. Table 1 provides a brief comparison between CT-COV19 and several other COVID-19 datasets, explaining the quantitative superiority of CT-COV19 to others.

TABLE I: A brief comparison between several publicly available COVID-19 CT-scan datasets.

| Dataset | #COVID-Images | # COVID-Patients | Availability |
|---|---|---|---|
| Ref. [19] | 20 | N.A. | Y |
| Ref. [24] | 100 | 60 | Y |
| Ref. [23] | 15k | 95 | Y |
| COVID-CT [46] | 349 | 216 | Y |
| SIRM COVID-19 Database [37] | 100 | 60 | Y |
| Ref. [19] | 35 | 144 | Y |
| CT-COV19 (this study) | 8500 | 500 | Y |

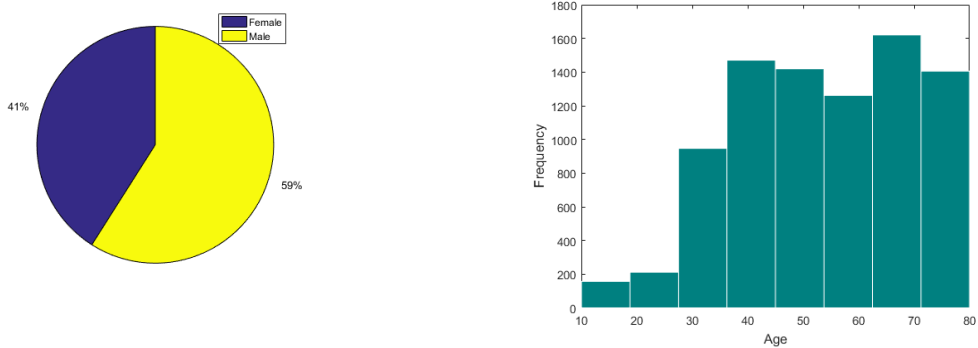

**Fig. 1:** Age and gender statistics in our proposed CT-COV19 dataset. **Left**: The gender ratio of COVID-19 patients. **Right**: The age distribution of patients with COVID-19

CT-COV19 consists of CT-scan images from three different classes including COVID-19, other pneumonia, and normal with the ratios of %61.5, %5.8, %32.7, respectively. Although the number of samples belonging to the other-pneumonia class is small, one can easily find plenty of such samples via the Internet. It is then worthy to stress that the main goal of CT-COV19 is providing valuable publicly available samples of COVID-19. The dataset is further randomly divided into train, validation, and test parts with ratios of %70, %10, and %20 respectively. Table 2 summarizes this division. The minimum and maximum of heights/widths are respectively $484 \times 484$ and $1024 \times 1024$. Additionally, the minimum resolution of the images is 150dpi and the bit depth is 24. Figure 2 provides several samples of this dataset.

### 2.2. Deep-CT-Net

This subsection introduces our proposed deep neural architecture (Deep-CT-Net), which can accurately classify CT-scan images into three classes including COVID-19, other kinds of pneumonia, and normal CT-images of lungs. Figure 3 depicts the workflow of our proposed Deep-CT-Net. As it is shown, Deep-CT-Net consists of three main parts. The first part applies Densnet-121 [14] as a backbone, extracting high-level features from the input raw CT-images.

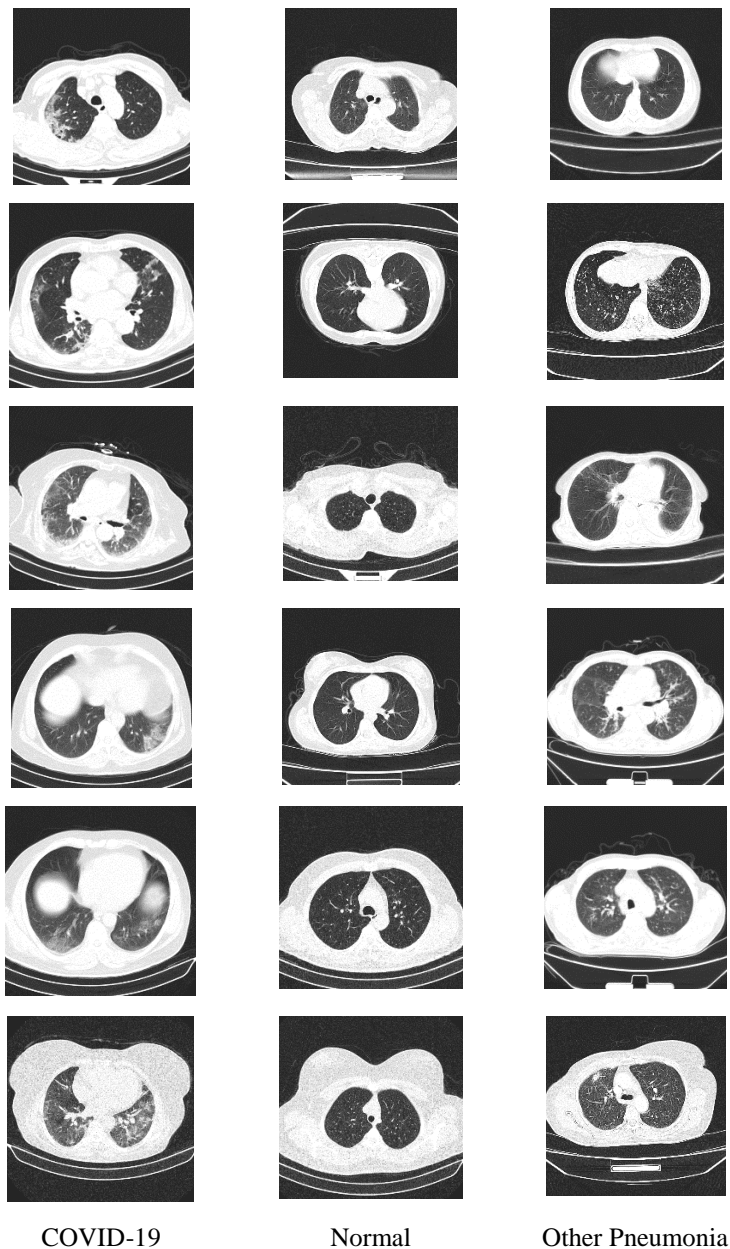

COVID-19      Normal      Other Pneumonia

**Fig. 2:** Several different instances of the proposed dataset, i.e., CT-COV19: each column provides several examples of each class presented in the dataset.

The second part performs a pyramid attention layer [20] over the extracted high-level features to maximize pixel-wise feature-extraction, allowing the model to detect COVID-19 even during the first days of infection. A batch normalization layer, which standardizes the input to the next layer, is then used to avoid internal covariance shifts [16] and have a smoother objective function [34].

The last part, finally, flattens the output of the previous part to be fed into fully-connected layers for prediction. The backward pass then updates the weight parameters of all three components using Adam optimizer with a learning rate of 1e-5 to minimize a binary-cross-entropy loss.

**TABLE II:** Train, validation and test splits in CT-COV19 dataset. The numbers denote the number of CT-scan images in each class (COVID-19 & Other pneumonia & Normal) and each split (train & validation & test.

| Data-part | COVID-19 | Other pneumonia | Normal | Total |
|---|---|---|---|---|
| Train | 6120 | 543 | 3060 | 9723 |
| Validation | 680 | 60 | 340 | 1080 |
| Test | 1700 | 151 | 851 | 2702 |
| Total | 8500 | 754 | 4251 | 13505 |

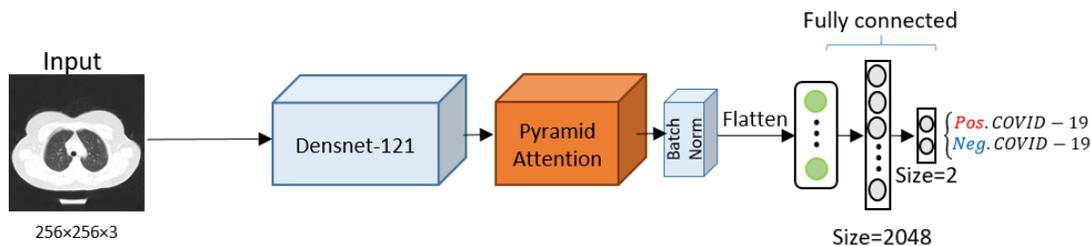

**Fig. 3:** (Deep-CT-Net) A graphical overview of the proposed fully supervised deep model for screening COVID-19 based on CT-scan data. More information about the network's components are provided in APPENDIX A.

### 2.3. Deep-CXR-Net

This subsection describes the Deep-CXR-Net, which is a transfer learning extension of Deep-CT-Net for screening COVID-19 based on CXR-images of lungs. The choice of transfer learning allows the Deep-CXR-Net to learn from a small set of labeled CXR COVID-19 images and provides a high-level of generalization ability. More accurately, the proposed Deep-CXR-Net, depicted in Figure 4, consists of three main parts, where the first two parts are independent pre-trained models respectively on two almost big datasets of non-COVID diseases in lungs, i.e., CheXpert [17] and Kaggle-Pneumonia [25]. The last part, i.e., Part 3, is another deep network whose linear layer concatenates extracted high-level features of all three parts. This concatenation is the point where the idea of transfer learning comes to play. More precisely, the additional concatenated features, which are provided by Parts 1 and 2, compensate for the lack of enough CXR training-samples whose labels are COVID-19 and provides a high-level of generalization at the testing time. We use ieee8023 [8] as a COVID-labeled dataset to train the parameters of the third part. Additionally, a certain number of data augmentation techniques, such as rotation and translation, has also been applied in the learning phase. As Figure 4 shows, the backbone applied in all three parts is DensNet-121. Similar to Deep-CT-Net, the proposed Deep-CXR-Net uses a pyramidal attention layer [20] to provide pixel-wise attention on high-level features, enabling the whole model to effectively detect COVID-19 cases even when there are small clues of the disease in lungs, i.e., detecting not only potential high-risk patients but also low-risk ones.

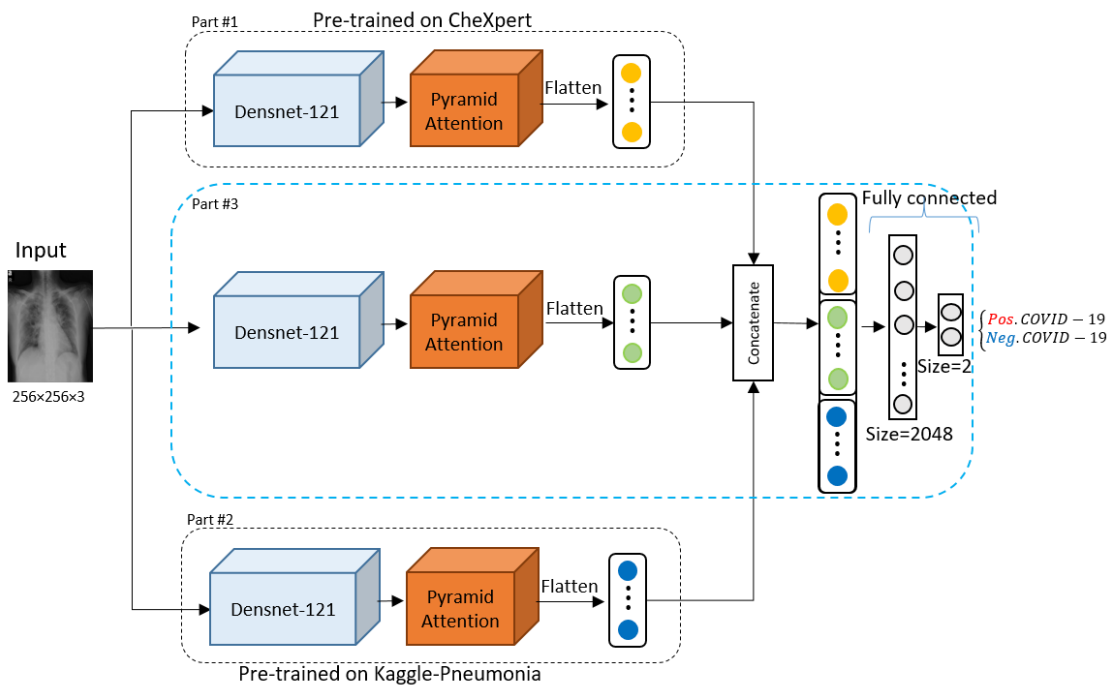

**Fig. 4:** (Deep-CXR-Net) A graphical overview of the proposed deep neural model for screening COVID-19 based on CXR data. Overall, Deep-CXR-Net consists of two pre-trained parts, extracting extra features to be concatenated with the features generated by the second part. The concatenated features are then passed thought fully connected layers to predict the label. More information about the network's components are provided in APPENDIX A.

## 3. PERFORMANCES OF DIAGNOSIS

We report the obtained results from the proposed methods and compare them with other related methods on our proposed CT-COV19 as well as several benchmark datasets. The obtained results are evaluated through several popular classification metrics.

### 3.1. The results of Deep-CT-Net

This subsection reports the obtained results from Deep-CT-Net through several experiments and provides a brief comparison. In the first experiment, we use our proposed dataset, i.e., CT-COV19, to evaluate and compare the screening results of the Deep-CT-Net with several deep network models [5, 13, 41, 44]. As shown in Table 3, our proposed model achieves an area under curve (AUC) of %88.6, which is superior to other related methods. Also, Figure 5 compares the obtained receiver operating characteristic (ROC) curves for each method, illustrating the superiority of the proposed Deep-CT-Net.

To have a better view over the proposed Deep-CT-Net, Figure 6 depicts Class Activation Mapping (CAM) [35] for a test sample of COVID-19, visualizing the attention regions of the lung. As shown in this figure, the attention regions detected by our proposed Deep-CT-Net are exactly related to COVID-19, i.e., the highlighted regions in red. However, most of the previous methods are not as accurate as Deep-CT-Net in their CAM visualization results. More precisely, their CAM results simply highlight a large proportion of lungs as their attention regions.

**TABLE III:** The obtained results on the proposed CT-COV19 dataset.

| Method | Precision | Recall | F-measure | AUC |
|---|---|---|---|---|
| COVNET [44] | 0.750 | 0.810 | 0.779 | 0.842 |
| DL-system [41] | 0.769 | 0.750 | 0.759 | 0.804 |
| Hybrid-3D [13] | 0.808 | 0.797 | 0.802 | 0.843 |
| Unet++[5] | 0.724 | 0.746 | 0.735 | 0.826 |
| **Deep-CT-Net** (this study) | 0.720 | 0.858 | 0.783 | 0.886 |

To assess the quality of the CT-COV19 dataset as well as our proposed CT-model, we applied Deep-CT-Net trained on our proposed CT-COV19 to the test set of COVID-CT dataset [46] without applying any fine-tuning or post-processing steps. It is worthy to note that COVID-CT actually consists of extracted CT-images of lungs from scientific documents and therefore does not provide high-quality images. Apart from that, images in COVID-CT are not in the same shapes. To clear this, Figure 7 depicts several images of this dataset. As it is shown, CT-images come with different types and illuminations. Table 4 reports the obtained results of Deep-CT-Net on this dataset and provides a comparison with the baseline methods reported in [46]. The results show that the proposed Deep-CT-Net trained on our proposed CT-COV19 could be generalized to a new set of unseen CT-scan images, i.e., the test set of COVID-CT.

Additionally, we evaluate the proposed Deep-CT-Net model on COVID-CT dataset and compare the obtained results with several state-of-the-art methods. Table 5 shows that the performance of Deep-CT-Net is superior to state-of-the-art models.

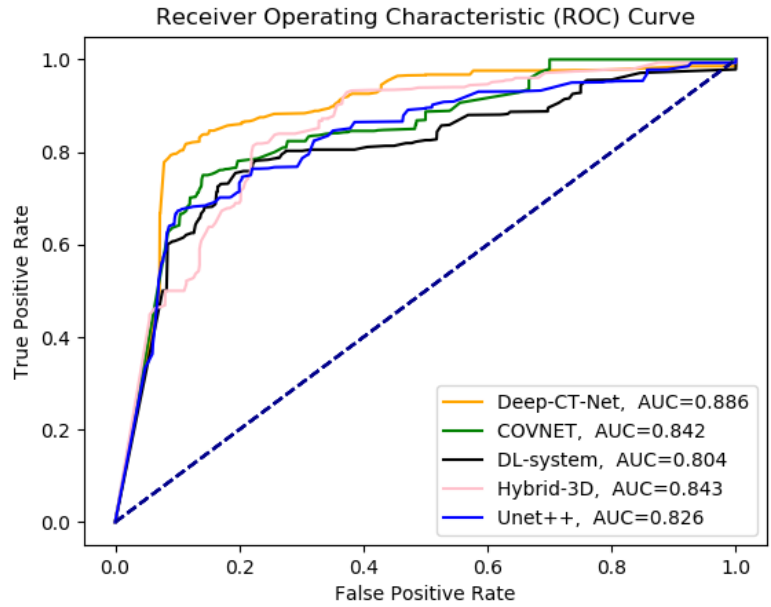

**Fig. 5:** The obtained receiver operating characteristic (ROC) curves over the proposed CT-COV19 dataset.

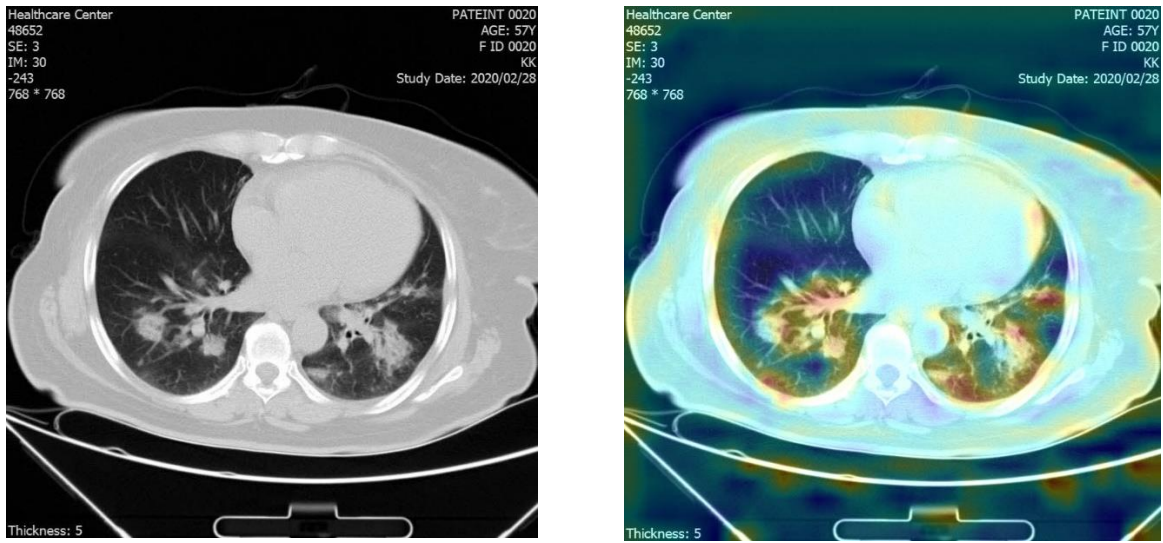

**Fig. 6:** Plotting the results of Class Activation Mapping (CAM) for a CT-image of COVID-19 (left): The CAM result (right) highlights the class-specific discriminative regions. The highlighted areas inside of the chest are discriminative regions.

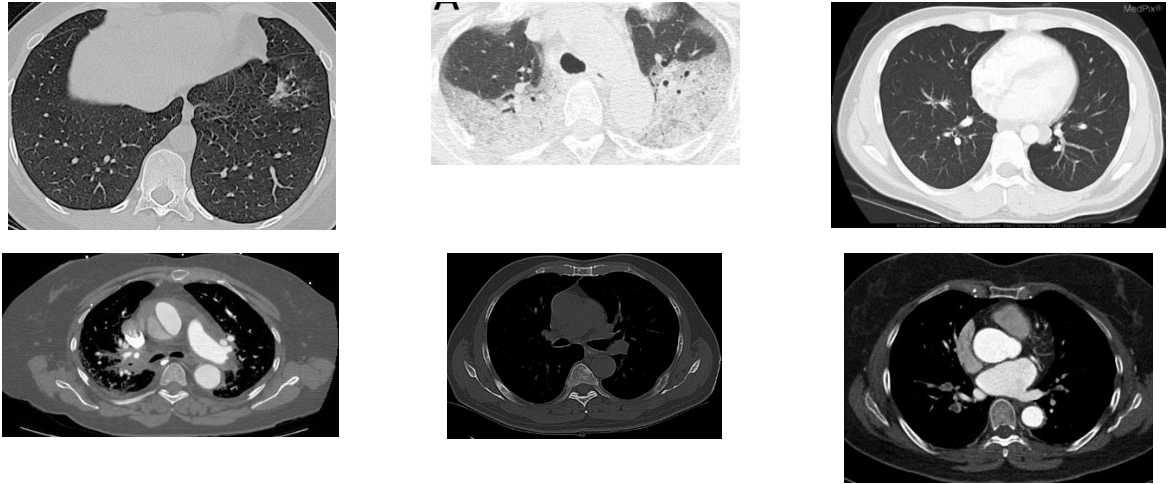

**Fig. 7:** Several instances of COVID-CT dataset [46]: The images come with diverse illumination, format, and shapes.

**TABLE IV:** The obtained results on COVID-CT dataset [46]. The proposed Deep-CT-Net was trained on our proposed CT-COV19 dataset and tested on COVID-CT without any fine-tuning. The results show the generalization ability of the proposed method.

| Method | F-measure | AUC | Accuracy |
|---|---|---|---|
| DenseNet-169 | 0.760 | 0.901 | 0.795 |
| ResNet-50 | 0.746 | 0.864 | 0.774 |
| **Deep-CT-Net** (trained on CT-COV19) | 0.801 | 0.920 | 0.860 |

### 3.2. The results of Deep-CXR-Net

The obtained results of Deep-CXR-Net are reported in this subsection. To have a reliable comparison, we randomly divided ieee8023 [8] to train and test sets with the proportions of %70 and %30 respectively. Then, we added 100 more CXR images of other Pneumonia labels to the test set. This set of additional test data is helpful for evaluating the performance of each method in terms of false-positive rate. Table 6 compares the obtained screening results of Deep-CXR-Net to other related methods [11, 12, 45]. It is important to note that the same experimental settings have been used for other methods. Overall, the proposed method performs significantly better than the others, while other methods have higher rates of false-positive in their predictions, i.e., they tend to wrongly predict samples of other pneumonia as COVID-19. Moreover, Figure 8 reports the obtained ROC curve for each method, illustrating the superiority of our proposed Deep-CXR-Net. Additionally, Figure 9 depicts Class Activation Map for a CXR-sample of COVID-19. The dotted areas are those regions of the lung that play an essential role in classifying this sample as COVID-19.

TABLE V: Comparison of the proposed Deep-CT-Net on the COVIDCT dataset [46]. The proposed Deep-CT-Net was learned on the train set of the COVID-CT dataset [46] dataset and evaluated on its test set.

| Method | Precision | Recall | F-measure | AUC |
|---|---|---|---|---|
| xDNN [2] | 0.897 | 0.886 | 0.892 | 0.886 |
| Baseline [46] | 0.970 | 0.762 | 0.853 | 0.824 |
| Ref. [29] | 0.817 | 0.850 | 0.833 | N.A. |
| **Deep-CT-Net** (this study) | 0.884 | 0.905 | 0.894 | 0.899 |

TABLE VI: The obtained results over CXR dataset (ieee8023).

| Method | Precision | Recall | F-measure | AUC | Accuracy |
|---|---|---|---|---|---|
| CAAD [45] | 0.943 | 0.915 | 0.924 | 0.920 | 0.930 |
| COVID-ResNet [11] | 0.926 | 0.916 | 0.920 | 0.916 | 0.916 |
| Ref. [12] | 0.898 | 0.894 | 0.896 | 0.889 | 0.889 |
| **Deep-CXR-Net** (this study) | 0.987 | 0.982 | 0.984 | 0.983 | 0.983 |

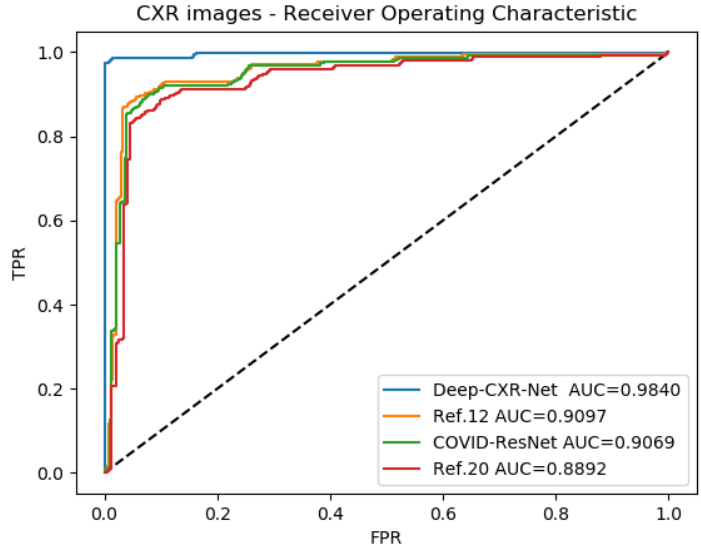

**Fig. 8:** The obtained ROC curves for ieee8023 dataset.

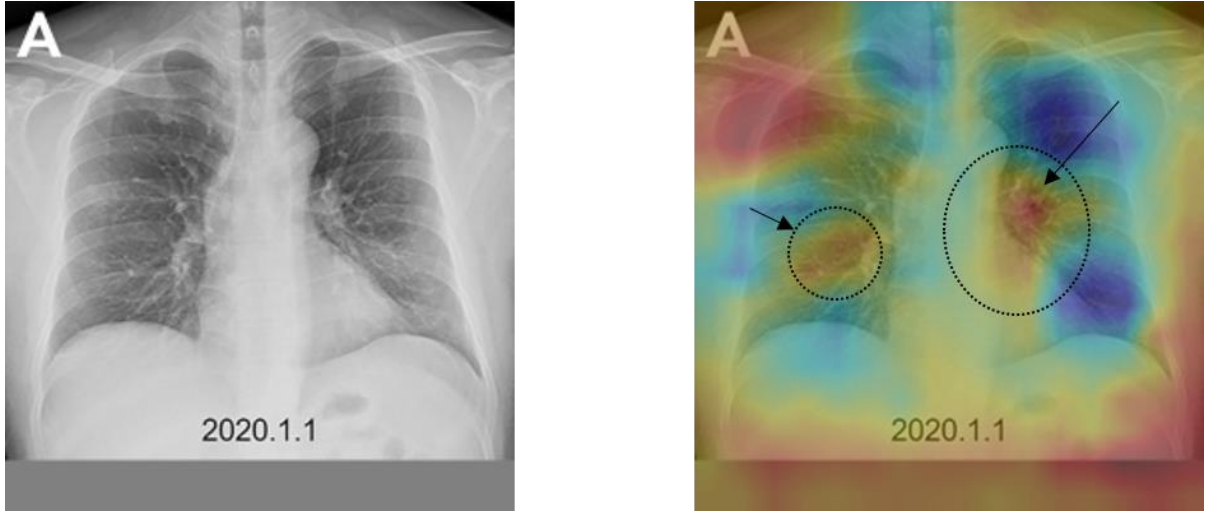

**Fig. 9:** Plotting the results of Class Activation Mapping (CAM) for an input CXR image (left): The CAM result (right) highlights the class-specific discriminative regions. All images belong to positive COVID-19 cases. The dotted regions inside of the chests are discriminative areas.

### 3.3. DISCUSSION

This work is significant from three main perspectives. First, we built a publicly available dataset of CT-images of COVID-19 that is large and diverse enough to train reliable models and, therefore, could be considered for training and evaluating future research. Second, the proposed Deep-CT-Net is able to accurately extract pixel-wise information and therefore very accurate to detect COVID-19. Third, the proposed CXR-COV-Net, which is the CXR extension of Deep-CT-Net, is able to be trained on small-sized CXR datasets and efficiently compensate for the lack of enough CXR data of COVID-19.

The proposed dataset could be considered as a general and diverse set of CT images for screening COVID-19, covering a broad range of COVID-19 patterns in the lungs. To investigate this claim, Deep-CT-Net has been trained on CT-COVID19 and tested on a completely new set of data, i.e., COVID-CT [46], without performing any fine-tuning step on this dataset. The obtained results are reported in Table 4, showing a significant performance compared to the baseline models.

To further evaluate the performance as well the generalizability of the proposed CT model, i.e., Deep-CT-Net, on another set of data, we apply the Deep-CT-Net to COVID-CT dataset. This dataset has a small number of samples with a lower image quality compared to CT-COVID19. Table 5 describes the obtained results and provides a comparison to several state-of-the-art models. As the table describes, Deep-CT-Net achieves a higher value of sensitivity, i.e., 0.905 compared to others. This observation certifies that the pyramid attention layer embedded in Deep-CT-Net is able to successfully detect a diverse range of infection patterns caused by COVID-19. Additionally, although the results show that the precision of Deep-CT-Net is lower than two other models, its overall score (F-measure) is better than others.

One of the main advantages of the proposed models is their efficiency in terms of computational complexities, allowing them to be used in ordinary computing systems of hospitals. More accurately, the size of models is not huge, particularly for the Deep-CT-Net, and also there are no complicated pre-processing steps. For instance, [13, 18] are based on a lung segmentation step, imposing more computational cost during the test phase and limiting the final screening results to the segmentation accuracy.

In conclusion, this study proposed a valuable set of COVID-19 CT images, allowing researchers to train more general deep models. We further proposed a baseline deep CT model, benefiting from a pyramidal attention layer that helps to extract pixel-wise features of COVID-19. Moreover, we extended our CT model for the case CXR images of lungs using a transform learning strategy, allowing the extended model to train on a small number of samples. To validate the generality of CT-COV19 as well as the proposed models, several different experiments are conducted.

As for future work, we plan to design a decoder for extending the current models to accurately segment the infected parts of the lungs with COVID-19, resulting in discovering biomarkers for COVID-19. This result could further be used to categorize the infection patterns of COVID-19, providing valuable sources of data to reveal the unknown aspects of this virus and eventually help in medical prescriptions. Additionally, Deep-CT-Net provides a baseline result over the proposed CT-COV19 dataset, and there is still room for developing more accurate deep models.

## 4. METHODS
### 4.1. Data Pre-processing

The following steps have been done for the input data in both proposed models, i.e., Deep-CT-Net and Deep-CXR-Net. The data pre-processing stage consists of five consecutive steps. The first step is applying random data augmentation techniques including a random rotation in a degree of [-15,15], and a random translation in a range of [-0.05,0.05]. The second step is performing a histogram equalization, which adjusts the contrast of a CT-scan image by modifying the intensity distribution of the histogram. The third step is fixing the aspect ratio of images (or simply image resizing) to a size of 256×256×3. Afterward, a Gaussian blur filtering with a window-size of 3 is used for the purpose of image smoothing. Finally, images are normalized by subtracting from their mean and divided by their standard deviation.

### 4.2. Network Architecture

As shown in Figure 3, the architecture of Deep-CT-Net has three components. The first one is DensNet-121, which consumes input CT images and returns extracted high-level features. The second component is a pyramidal attention layer, operating on the output of the first component and extract pixel-wise information from the data. Finally, the third component, which is a fully connected layer, consumes the output of the second component and predicts the binary labels. The backward pass updates the weight parameters of all three components using Adam optimizer with a learning rate of 1e-5 to minimize the binary-cross-entropy loss.

Figure 4 shows the architecture of the Deep-CXR-Net. It has three main parts, where the third-part consumes the outputs of the first two parts in its fully-connected layers. As shown in Figure 4, the main network structures in all three parts are the same. The first two parts are pre-trained networks on two popular CXR datasets, i.e., CheXpert and Kaggle-Pneumonia, respectively. We consider these parts as two black-box functions whose inputs are CXR images and outputs are vectors of high-level features. More precisely, for a given CXR image, the output of the first part is a six-dimensional vector whose entries are the likelihood of six certain lung diseases. Besides, the output of the second part is a two-dimensional vector, representing the likelihood of having pneumonia or not. Afterward, the third-part considers these outputs as additional extracted features and concatenates them to form a feature vector. These additional features compensate for the scarcity of CXR data in COVID-19, increasing the generalization ability of the Deep-CXR-Net. Finally, the feature vector from the concatenation layer is passed through two fully-connected layers to perform binary prediction $y \in \{0,1\}$. The backward pass updates the weight parameters of the third part using Adam optimizer with a learning rate of 1e-5 to minimize the binary-cross-entropy loss. Compared with other CXR-based deep models, we found that the choice of using additional features results in much better performances for unseen CXR data and significantly reduces the rate of false positive.

# APPENDIX A

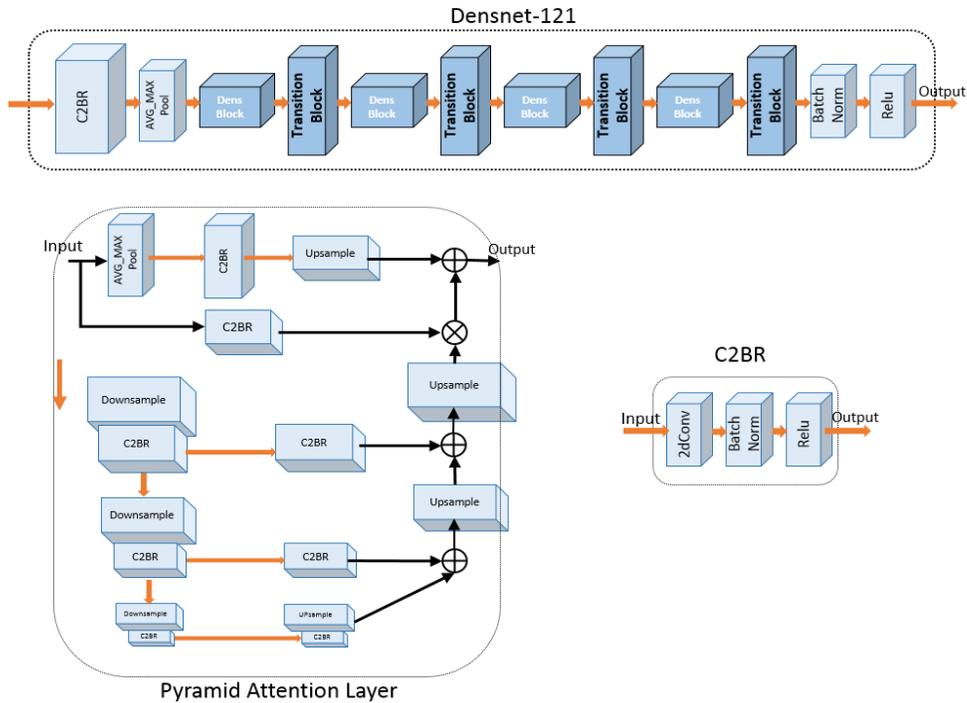

**Fig. 10:** A graphical explanation of the components depicted in Figure 3.

## ACKNOWLEDGEMENTS

This study was not funded by any national or private institution. We also would like to acknowledge the radiologists and scientists helping us in this study.

## AUTHOR CONTRIBUTIONS

M.D. conceived the main ideas, designed the models, analyzed the results (lead), and wrote the manuscript. A.H. supervised the research. H.R. co-supervised the research, reviewed the work, and guide (supporting) the research. AR. R. analyzed the results (supporting) and contributed to prepare the CT dataset. S. Dialameh confirmed the results and contributed in data correctness.

## COMPETING INTERESTS

The authors declare no competing interests.

## DATA AVAILABILITY

Our proposed dataset, i.e., CT-COV19, is publicly reachable via this link:
**https://github.com/m2dgithub/CT-COV19.git**.